%% file: dihadron_JHEP.tex
\documentclass[notoc]{JHEP3}

\usepackage{graphicx}% Include figure files
\usepackage{dcolumn}% Align table columns on decimal point
\usepackage{bm}% bold math
\usepackage{rotating}
\usepackage{amsmath,amsfonts}

\long\def\symbolfootnote[#1]#2{\begingroup%
\def\thefootnote{\fnsymbol{footnote}}\footnote[#1]{#2}\endgroup}

\def\hermesauthor[#1]#2{{#2}$^{\, #1}$}
\def\hermesinstitute[#1]#2{$^{#1\,}$ {#2}\\}

\newcommand{\av}[1]{\langle #1 \rangle}
\newcommand{\newangle}{\sphericalangle}
\newcommand{\mb}{\boldsymbol}

\def\mpipi{\ensuremath{M_{\pi\pi}}}
	
\def\hermes{{\sc Hermes}}
\def\compass{{\sc Compass}}
\def\belle{{\sc Belle}}

\def\rhic{{\sc Rhic}}

\def\desy{{\sc Desy}}
\def\hera{{\sc Hera}}

\def\pythia6{{\sc Pythia6}}
\def\geant3{{\sc Geant3}}

\newcommand{\de}{{\rm\,d}}

\title{Evidence for a Transverse Single-Spin Asymmetry in Leptoproduction
of ${\pi^+\pi^-}$ Pairs}

\author{\hermes\ Collaboration \\ HERMES -- DESY, Notkestra\ss e 85, D-22607 Hamburg}

\abstract{A single-spin asymmetry was measured in 
the azimuthal distribution of $\pi^+\pi^-$ pairs produced in semi-inclusive 
deep-inelastic scattering on a transversely polarized hydrogen target. 
For the first time, evidence is found for a correlation between the 
transverse target polarization and the 
azimuthal orientation of the plane containing the two pions.
The corresponding single-spin asymmetry is expected to be related to the product of the little-known
quark transversity 
distribution function and an unknown naive-T-odd chiral-odd dihadron
fragmentation function. }

\preprint{arXiv:0803.2367}

\begin{document}

Three fundamental parton distribution functions describe the
structure of the nucleon at leading twist: the unpolarized distribution, the
helicity distribution, and the transversity distribution. 
Transversity describes the distribution of transversely polarized quarks 
in a  nucleon with polarization transverse to the direction of the hard probe and is the
most difficult one to measure. (For a review  see
Ref.~\cite{Barone:2001sp}.) Unlike the other two, 
it is inaccessible in inclusive deep-inelastic scattering (DIS). 
A class of observables sensitive to the
transversity distribution is that of single-spin asymmetries in
semi-inclusive DIS on a transversely polarized target. 

In general, 
single-spin asymmetries are related to mixed products of the type 
$\boldsymbol{S}\cdot(\boldsymbol{P}_1\times \boldsymbol{P}_2)$, 
where $\boldsymbol{S}$ is a spin vector (typically the spin of the target or
of the quark), and $\boldsymbol{P}_1$ and $\boldsymbol{P}_2$ are two noncollinear
momenta.
Single-spin asymmetries are odd under naive time reversal (naive-T-odd), i.e., time reversal
without the interchange of initial and final states~\cite{DeRujula:1971te}. 
Single-spin
asymmetries are sensitive to physics at the amplitude level, 
as they can arise only from the interference between two scattering
amplitudes with different phases. Because of the structure of the
mixed product, single-spin asymmetries require an interplay between a spin
and an orbital angular momentum.

Azimuthal single-spin asymmetries in {\it single}-hadron production
in semi-inclusive DIS ($ep \rightarrow e'hX$)
on a transversely polarized target were recently measured
by the \hermes\ collaboration for charged 
pions~\cite{Airapetian:2004tw} and
by the \compass\ collaboration for unidentified charged 
hadrons~\cite{Alexakhin:2005iw,Ageev:2006da}.
For these observables, the orientation of the target transverse
polarization influences the distribution of hadrons in the azimuthal angle
around the virtual-photon direction through, e.g.,  the so-called
Collins~\cite{Collins:1993kk} and Sivers~\cite{Sivers:1990cc} mechanisms. 
In particular, 
the Collins asymmetry is sensitive to the transversity distribution. At the
partonic level, this asymmetry arises from  the process in which initially 
a transversely polarized quark in the
transversely polarized target absorbs the virtual photon. The orientation 
of the transverse polarization of the quark changes in a manner calculable using QED. 
In the subsequent hadronization of the quark, 
the direction of the momentum of the detected hadron can be
related to the direction of the spin of the quark 
via the mixed product
$\boldsymbol{S}_q\cdot(\boldsymbol{p}_q\times \boldsymbol{P}_h)$, where
$\boldsymbol{p}_q$ is the momentum of the struck quark,
$\boldsymbol{S}_q$ its spin and $\boldsymbol{P}_h$ is the momentum of the
detected hadron.
If such a correlation exists, the hadron has a preference to move to a
specific side with respect to the quark spin and the direction of its momentum.
The effect vanishes when 
integrating over the component of the detected
hadron's momentum transverse to the momentum of the
fragmenting quark. From a formal point of view, 
despite the complications due to the presence of 
transverse momentum,
factorization proofs~\cite{Ji:2004xq,Collins:2004nx}
allow the interpretation of the Collins asymmetry in terms of a
convolution in quark transverse-momentum space of the transversity
distribution with a universal naive-T-odd fragmentation function, the Collins function,
which can be considered as an
analyzer of the fragmenting quark's transverse polarization. 
This function can be measured in
other processes, e.g., in $e^+ e^-$ collisions, and can then be used to extract the
transversity distribution from the above asymmetries~\cite{Anselmino:2007fs}.
The only existing data that have been used to isolate transversity are from
such measurements of single-spin asymmetries of single hadrons in semi-inclusive DIS.

By the early 1990s it had already been pointed out that single-spin asymmetries in
semi-inclusive {\it dihadron}\footnote{The two hadrons, i.e., $h_1$ and $h_2$, 
have to be of different hadron types.} production 
($ep \rightarrow e'h_1h_2X$) on a transversely polarized target
could also be sensitive to transversity~\cite{Efremov:1992pe,Collins:1994kq}, 
thereby providing an independent experimental constraint.
The underlying mechanism differs from the Collins mechanism in that  the transverse spin of the fragmenting quark is transferred to the relative orbital angular momentum of the hadron pair.
Consequently, this mechanism does not require transverse momentum of the hadron pair. 

Dihadron fragmentation functions were introduced in
Ref.~\cite{Konishi:1978yx}. 
Polarized dihadron fragmentation functions were studied in
Refs.~\cite{Collins:1994kq,Collins:1994ax,Jaffe:1998hf,Artru:1996zu}. 
They are related to the concept of
jet-handedness~\cite{Efremov:1992pe,Stratmann:1992gu}, as explained in 
Ref.~\cite{Boer:2003ya}. The decomposition of the cross section in terms of
quark-distribution and dihadron-fragmentation functions was carried out to
leading twist (twist-2) in  Ref.~\cite{Bianconi:1999cd} and to twist-3 in
Ref.~\cite{Bacchetta:2003vn}.  
Polarized $\rho^0$ fragmentation
functions~\cite{Efremov:1982sh,Ji:1994vw,Anselmino:1999cg,Bacchetta:2000jk}
are ($p$-wave) components of dihadron fragmentation functions, as
reflected in the angular distribution of the decay products of the $\rho^0$ meson.

Little experimental information exists on
the multidimensional kinematic dependence of dihadron fragmentation functions.
Invariant-mass spectra of hadron pairs were measured in a number of experiments, 
some of which studied semi-inclusive
DIS~\cite{Cohen:1982zg,Aubert:1983un,Arneodo:1986tc}.
Dihadron fragmentation functions have recently been studied in a nuclear
environment \cite{Airapetian:2005yh}, as they might be relevant to
the phenomenon of jet quenching in heavy-ion
physics~\cite{Majumder:2004wh}. 
Vector-meson polarization was analyzed in $e^+ e^-$ and $pp$
collisions~\cite{Anselmino:1999cg,Abreu:1997wd,Abbiendi:1999bz,Xu:2001hz,Xu:2003fq}.   
However, these data were not interpreted in terms of dihadron fragmentation
functions. Finally, studies of longitudinal jet-handedness gave
results consistent with zero~\cite{Abe:1994bk}.

\FIGURE{
\includegraphics[width=10.8cm]{dihadronplanes.epsi}
\caption{Depiction of the azimuthal angles $\phi_{R \perp}$ of the dihadron
	and  $\phi_S$ of the component \(\boldsymbol{S_T}\) of the target-polarization 
	transverse to both the virtual-photon and target-nucleon momenta 
	$\boldsymbol{q}$ and $\boldsymbol{P}$, respectively. Both angles are 
	evaluated in the virtual-photon-nucleon center-of-momentum frame. 
	Explicitly,  
	$\phi_{R \perp} \equiv \frac{(\boldsymbol{q} \times \boldsymbol{k})
  	\cdot\boldsymbol{R_T}}{|(\boldsymbol{q} \times \boldsymbol{k})
  	\cdot\boldsymbol{R_T}|}  \arccos {\frac{
  	(\boldsymbol{q} \times \boldsymbol{k}) \cdot
  	(\boldsymbol{q}\times\boldsymbol{R_T})}{|\boldsymbol{q} \times
  	\boldsymbol{k}| |\boldsymbol{q}\times\boldsymbol{R_T}|}}$   
	and $\phi_S \equiv
  	\frac{(\boldsymbol{q}\times\boldsymbol{k}) \cdot
  	\boldsymbol{S_T}}{|(\boldsymbol{q}\times\boldsymbol{k}) \cdot
  	\boldsymbol{S_T}|} \arccos {\frac{(\boldsymbol{q} \times
  	\boldsymbol{k}) \cdot (\boldsymbol{q} \times \boldsymbol{S_T})}{|
  	\boldsymbol{q}\times\boldsymbol{k}||\boldsymbol{q} \times
  	\boldsymbol{S_T}|}}$.  
	Here, 
	$\boldsymbol{R_T} = \boldsymbol{R} - (\boldsymbol{R}\cdot
  	\hat{\boldsymbol{P_h}})\hat{\boldsymbol{P_h}}$, with $\boldsymbol{R} \equiv
  	(\boldsymbol{P_{\pi^+}} - \boldsymbol{P_{\pi^-}})/2$, $\boldsymbol{P_h}
  	\equiv \boldsymbol{P_{\pi^+}} + \boldsymbol{P_{\pi^-}}$, 
  	and 
	$\hat{\boldsymbol{P_h}}\equiv \boldsymbol{P_h}/ \mid\boldsymbol{P_h}\mid$,
  	thus $R_T$ is the component of $P_{\pi^+}$ orthogonal to $P_h$, and 
	$\phi_{R\perp}$ is the azimuthal angle of $R_T$ about the virtual-photon direction.
	The dotted lines indicate how vectors are projected onto planes.  
	The short dotted line is parallel to the direction of the virtual photon.
	Also included is a description of the polar angle $\theta$, which is evaluated 
	in the center-of-momentum frame of the pion pair.}
\label{fig:azimuthal_angle}
}

Denoting $2\bm{R}$ as the difference of the momenta of the two hadrons $h_1$ and $h_2$,
the hadronization of a transversely polarized quark into the hadron pair can depend on  
the mixed product  $\boldsymbol{S}_q\cdot(\boldsymbol{p}_q\times \boldsymbol{R})$.
This would imply a preference of $h_1$ to go to a
specific side with respect to the spin and the momentum direction
of the quark, while $h_2$ would go to the opposite side.
This preference is revealed in the
cross section through a dependence on the angle $\phi_{R \perp}$, the azimuthal
angle of $\bm{R_T}$, the component of $\bm{R}$ transverse to $\bm{P_h}$ 
(see Fig.~\ref{fig:azimuthal_angle} for the case of $\pi^+\pi^-$ pairs). 
Here, $\bm{P_h}$ is the sum of the momenta of the two hadrons.
Since $\phi_{R \perp}$ is the azimuthal orientation of the {\it relative} transverse momentum of
the two hadrons, the correlation described above remains present even if
the cross section is integrated over the transverse component $\bm{P_{h\perp}}$
of $\bm{P_h}$.
The benefits of integrating over $\bm{P_{h\perp}}$ are the following: 
i)~issues related to factorization are simpler [34], 
ii) the evolution equations for the fragmentation functions involved are known  
\cite{deFlorian:2003cg,Ceccopieri:2007ip}, 
iii) distribution and fragmentation functions appear in a simple product instead
of a convolution integral over transverse momentum.

This paper reports a measurement of an azimuthal Fourier amplitude of a single-spin asymmetry in
semi-inclusive \(\pi^+\pi^-\) production on a transversely polarized hydrogen target, resulting in
the first evidence of a naive-T-odd chiral-odd dihadron fragmentation function that can
provide access to transversity.
It is related to the product of the twist-two chiral-odd transversity distribution
$h_1^q$ (also called $\delta q$) for quark flavor $q$ and the 
twist-two chiral-odd naive-T-odd dihadron fragmentation function
$H_{1,q}^{\newangle}$~\cite{Bacchetta:2003vn,Radici:2001na}.\footnote{The 
superscript ${\newangle}$ indicates that the fragmentation function does not survive 
integration over the relative momentum of the hadron pair.}
There are no contributions to this amplitude at subleading twist (i.e., twist-3).
Among the various contributions to the fragmentation function $H_{1,q}^{\newangle}$ 
are the interference $H_{1,q}^{\newangle,sp}$ between the $s$- and $p$-wave 
components of the \(\pi^+\pi^-\) pair  and the interference $H_{1,q}^{\newangle,pp}$
between two $p$-waves. 
In some of the literature, such functions have therefore been called {\it
  interference} fragmentation functions~\cite{Jaffe:1998hf}, even though in
general interference between different
amplitudes is required by {\it all} naive-T-odd functions. 
In this paper the focus is on the $sp$-interference, since it has
received the most theoretical attention. 
In particular, in Ref.~\cite{Jaffe:1998hf} $H_{1,q}^{\newangle,sp}$
was predicted to change sign at a very specific 
value of the  invariant mass $M_{\pi\pi}$ of the \(\pi^+\pi^-\) pair, 
close to the mass of the $\rho^0$ meson. However, other
models~\cite{Radici:2001na,Bacchetta:2006un} predict
a completely different behavior.

The data presented here were recorded during the 2002-2005 running period
of the \hermes\ experiment, using the 27.6\,\,GeV  positron or electron beam and 
a transversely polarized hydrogen gas target internal to the \hera\ storage ring at
\desy. 
The open-ended target cell was fed by an atomic-beam source
\cite{Nass:2003mk} based on Stern-Gerlach separation combined with
transitions of hydrogen hyperfine states.
The nuclear polarization of the atoms was flipped at
1--3\,\,min.~time intervals, while both this polarization and the atomic
fraction inside the target cell were continuously measured
\cite{Airapetian:2004yf}. The average value of the transverse 
proton polarization $|{S_\perp}|$ was $ 0.74 \pm 0.06$.

Scattered leptons and coincident hadrons were detected by the
\hermes\ spectrometer \cite{Ackerstaff:1998av}. Its acceptance spanned the
scattering-angle range $40 < |\theta_y| < 140$\,\,mrad and $|\theta_x|
<170$\,\,mrad, corresponding to an almost full coverage in $\phi_S$ from 0 to $2 \pi$ with 
only small gaps at 1.40 $< \phi_S < 1.74$ rad and 4.54 $< \phi_S <$ 4.88 rad.
Leptons were identified  with an efficiency exceeding 98\%
and a hadron contamination of less than 1\% using an electromagnetic 
calorimeter, a transition-radiation detector, a preshower scintillation
counter, and a 
dual-radiator ring-imaging {\v C}erenkov detector \cite{Akopov:2000qi}, 
mainly used here also to identify 
charged pions with momentum $|\boldsymbol{P_\pi}| > 1$\,\,GeV.

Events were selected with the kinematic requirements $W^2 > 10$\,\,GeV$^2$,  
$0.1 < y < 0.85$, and $Q^2 > 1$\,\,GeV$^2$, where
$W$ is the invariant mass of the initial photon-nucleon system and $y = (P
\cdot q)/(P \cdot k)$, with $P$, $q$, and $k$ representing the four-momenta of
the target nucleon, the virtual photon, and the incident lepton, respectively. 
A constraint was placed on the missing mass: $M_X> 2 $\,\,GeV.
This avoids contributions from exclusive 
two-pion production, where factorization in distribution and fragmentation functions 
cannot be applied.
All possible combinations of detected $\pi^+\pi^-$ pairs were included for each
event, in contrast to keeping only the combination with the largest energy
fraction $z$, a choice for which fragmentation functions are not defined. Here, $z$ refers to 
the fraction of the energy $\nu$ of the virtual photon (in the target rest frame) that is
transferred to the pion pair, i.e., $z = (E_{\pi^+} + E_{\pi^-})/\nu =
z_{\pi^+} + z_{\pi^-} $.

%%%%%%%%%%%%%%  THEORY EXPRESSIONS %%%%%%%%%%%%%%%%

In semi-inclusive deep-inelastic scattering of an unpolarized ($U$) beam off
an unpolarized ($U$) target, the cross section $\sigma_{UU}$ for the production of pion pairs, 
integrated over the transverse momentum $P_{h\perp}$ of the pion pair,
is given, at leading twist and in leading order in $\alpha_s$ ($\alpha_s^0$), 
by \cite{Bacchetta:2002ux}
\begin{equation}
\frac{\de^7 \sigma_{UU}}{\de x \de y \de z \de \phi_S \de \phi_{R\perp} \de\!\cos \theta \de M_{\pi\pi}}
= \sum_q \frac{\alpha^2 e_q^2 }{2 \pi s x y^2} (1-y+\frac{y^2}{2}) f_1^q(x) D_{1,q}(z,M_{\pi \pi},\cos \theta),
\label{eq:sigmaUU}
\end{equation}
where $\alpha$ is the fine-structure constant, $x=Q^2/(2P\cdot q)$, the Mandelstam invariant $s=(P+k)^2$,
$f_1^q$ is the polarization-averaged quark distribution 
function and $D_{1,q}$ is a dihadron fragmentation function representing the number density of pion pairs produced from unpolarized quarks.  
The summation runs over the quark and antiquark flavors $q$  with charges $e_q$ in units of the elementary charge.
For an unpolarized beam and integrating over $P_{h\perp}$, 
the cross section difference $\sigma_{UT}$ of the polarized cross sections 
$\sigma_{U\uparrow}$ and $\sigma_{U\downarrow}$, where the target is in either of the two
corresponding opposite transverse ($T$) spin states $\uparrow\downarrow$, 
is given at leading twist and in leading order in $\alpha_s$ by \cite{Bacchetta:2002ux}
\begin{eqnarray}
%\begin{multline}
\nonumber 
\hspace*{-.8cm}& & \frac{\de^7 \sigma_{UT}}{\de x \de y \de z \de \phi_S \de \phi_{R\perp} \de\! \cos \theta \de M_{\pi\pi}} \equiv \frac{1}{2} \left( \de^7 \sigma_{U\uparrow}- \de^7 \sigma_{U\downarrow} \right) = \\
\hspace*{-.8cm}& & - |\boldsymbol{S_T}| \sum_q \frac{\alpha^2 e_q^2 }{2 \pi s x y^2} (1-y) 
                              \frac{1}{2}\sqrt{1-4 \frac{M_\pi^2}{M_{\pi\pi}^2}} 
\sin(\phi_{R \perp} + \phi_S)\sin\theta  \ h_1^q(x) H_{1,q}^{\newangle}(z,M_{\pi \pi},\cos \theta),
\label{eq:sigmaUT}
%\end{multline}
\end{eqnarray}
where $M_\pi$ is the pion mass and 
\(\boldsymbol{S_T}\) is the component of the target spin \(\boldsymbol{S}\) perpendicular to the 
virtual-photon direction. The azimuthal angle $\phi_S$  always refers to the spin direction, relative to the lepton-scattering plane, of the target ``$\uparrow$'' state.
Twist-3 contributions to the polarized and unpolarized cross sections appear with different azimuthal dependences~\cite{Bacchetta:2003vn}. 

Both dihadron fragmentation functions $D_{1,q}$ and $H_{1,q}^{\newangle}$ can be expanded in terms of Legendre functions of $\cos \theta$. Hence~\cite{Bacchetta:2002ux},
\begin{equation}
 D_{1,q}(z,M_{\pi \pi},\cos \theta) \simeq D_{1,q}(z,M_{\pi \pi}) +  
D_{1,q}^{sp}(z,M_{\pi \pi})  \cos \theta  +  
D_{1,q}^{pp}(z,M_{\pi \pi})\frac{1}{4} (3\cos^2 \theta - 1) 
\label{eq:D1expansion}
\end{equation}
and 
\begin{equation}
H_{1,q}^{\newangle}(z,M_{\pi \pi},\cos \theta) \simeq 
					       H_{1,q}^{\newangle, sp}(z,M_{\pi \pi}) + 
					H_{1,q}^{\newangle,pp}(z,M_{\pi \pi}) \cos \theta , 
\label{eq:H1expansion}
\end{equation}
where the Legendre expansions are truncated to include only the $s$- and $p$-wave
components, which is assumed to be a valid approximation 
in the range of the invariant mass $M_{\pi \pi}<1.5$~GeV~\cite{Bacchetta:2002ux}, 
which is typical of the present experiment.

In Refs.~\cite{Jaffe:1998hf,Radici:2001na,Bacchetta:2002ux}, it was proposed to measure 
$\sigma_{UU}$ and $\sigma_{UT}$ integrated over the angle $\theta$, 
which has the advantage that
in the resulting expression for these cross sections the only fragmentation
functions that appear are $D_{1,q}(z,\mpipi)$ and $H_{1,q}^{\newangle,sp}(z,\mpipi)$ 
(see Eqs.~(\ref{eq:sigmaUU}-\ref{eq:H1expansion})). 
However, this requires an experimental acceptance that is complete in $\theta$, which is
difficult to achieve, not only because of the geometrical acceptance
of the detector, but also because of the acceptance in the momentum of the detected
pions. As the momentum selection $|\boldsymbol{P_\pi}| > 1$\,\,GeV strongly
influences the $\theta$ distribution, the measured asymmetry must
be kept differential in $\theta$.

The single-spin asymmetry
\(A_{UT}\equiv\frac{1}{|\boldsymbol{S_T}|} \sigma_{UT}/\sigma_{UU}\)
contains  components of a simultaneous Fourier and Legendre expansion.
The amplitude $A_{UT}^{\sin(\phi_{R \perp} + \phi_S)\sin\theta}$ of the 
modulation of interest here,
which is related to the product of transversity and the fragmentation function
$H_1^{\newangle,sp}$, 
is defined as
%The amplitude $A_{UT}^{\sin(\phi_{R \perp} + \phi_S)\sin\theta}$ of the 
%single-spin asymmetry
%\(A_{UT}\equiv\frac{1}{|\boldsymbol{S_T}|} \sigma_{UT}/\sigma_{UU}\),  
%which is related to the product of transversity and the fragmentation function
%$H_1^{\newangle,sp}$, 
%is defined as
%
\begin{equation}
A_{UT}^{\sin(\phi_{R\perp} + \phi_S)\sin\theta} \equiv 
 \frac{2}{|\boldsymbol{S_T}|}  \frac{\int \!\! \de\! \cos\theta \de \phi_{R\perp} \de \phi_S \, \sin(\phi_{R\perp}+\phi_S)\de \sigma^7_{UT}/\sin\theta}{\int \!\! \de\! \cos\theta \de \phi_{R\perp} \de \phi_S \de \sigma^7_{UU}} .
\label{eq:moment_definition}
\end{equation}
Using Eqs.~(\ref{eq:sigmaUU}-\ref{eq:H1expansion}), it can be written as~\cite{Bacchetta:2002ux}
\begin{equation}
	A_{UT}^{\sin(\phi_{R\perp} + \phi_S)\sin\theta}  = 
	-\frac{(1-y)}{(1-y+\frac{y^2}{2})}
	\ \frac{1}{2} \sqrt{1-4\frac{M_\pi^2}{M_{\pi\pi}^2}} \ \ 
	\frac{\sum_q e_q^2 \,h_1^q(x)\,H_{1,q}^{\newangle,sp}(z,M_{\pi\pi})}{\sum_q 
	e_q^2 \,f_1^q(x)\,D_{1,q}(z,M_{\pi\pi})} .\label{eq:A_UT_theory}
\end{equation}
Due to the factor $e_q^2$, the amplitude is expected to 
be $up$-quark dominated.

%
%
%
%%%%%%%%       Extraction of Amplitudes      %%%%%%%%%%%%%
%
%
%

The results reported here are extracted from the single-spin asymmetry
\begin{equation}
A_{U\perp} (x,z,M_{\pi\pi}, \phi_{R \perp}, \phi_S, \theta)
\equiv 
\frac{1}{|{S_\perp}|}
\frac{N^{\uparrow} -  N^{\downarrow}}{N^{\uparrow} + N^{\downarrow}}
\mathrm{,}
\label{eq:asymmetry}
\end{equation}
where $N^{\uparrow (\downarrow)}$ is the luminosity-normalized number of semi-inclusive 
$\pi^+\pi^-$ pairs detected while the target is in the $\uparrow$($\downarrow$) spin state with 
polarization perpendicular to the \emph{incoming lepton beam} (rather than to the virtual-photon direction). 
The asymmetry is evaluated as a function of  $x$, $z$, $M_{\pi\pi}$, 
and the angles $\phi_{R \perp}$, $\phi_S$, and $\theta$, which are defined in 
Fig.~\ref{fig:azimuthal_angle}.\footnote{The definitions of the asymmetry and 
the angles are consistent with the ``Trento Conventions''~\cite{Bacchetta:2004jz}.}

A $\chi^2$ fit was performed, binned in $(\phi_{R\perp}+\phi_S)$ 
versus $\theta'\equiv ||\theta-\pi/2|-\pi/2|$, with a 
function of the form:
\begin{equation}
A_{U\perp} (\phi_{R \perp} + \phi_S,  \theta')
= \sin(\phi_{R \perp} + \phi_S)\frac{a \sin \theta'}{1 + b
\displaystyle{\frac{1}{4}}(3\cos^2 \theta' -1)}\mathrm{,}\label{eq:fit_function}
\end{equation}
where $a\equiv A_{U\perp}^{\sin(\phi_{R\perp} + \phi_S)\sin\theta} $ 
is a free parameter of the fit, while $b$ is varied to study 
the influence of the unknown
contribution $D_{1,q}^{pp}$ to the polarization-averaged 2-hadron cross section. 
The fit is evaluated as a function 
of $\theta' $, which corresponds to a symmetrization of the fit around $\theta = \pi/2$. 
This has the advantage that the contributions to $A_{U\perp}$ containing $D_{1,q}^{sp}$ and 
$H_{1,q}^{\newangle,pp}$ drop out (see, e.g., Eqs.~\eqref{eq:D1expansion} and 
\eqref{eq:H1expansion}), reducing the statistical uncertainty on 
$a \equiv A_{U\perp}^{\sin(\phi_{R\perp} + \phi_S)\sin\theta}$, the modulation amplitude of interest that
approximates $A_{UT}^{\sin(\phi_{R \perp} + \phi_S)\sin \theta}$ defined in 
Eq.~\eqref{eq:moment_definition}.

The value of the fit parameter $a$ depends on the value of  $b$. 
Therefore, a systematic uncertainty was assigned to the extracted value of $a$ by studying its
response to variation of $b$.  The
parameter $b$ was varied within its positivity limits, given by~\cite{Bacchetta:2002ux}
\begin{equation}
-\frac{3 D_{1,q}^{p}(z,\mpipi)}{2 D_{1,q}(z,\mpipi)}\le b \le \frac{3 D_{1,q}^{p}(z,\mpipi)}{D_{1,q}(z,\mpipi)},
\end{equation} 
where $D_{1,q}^{p}(z,\mpipi)$ indicates the pure $p$-wave component of the fragmentation functions $D_{1,q}(z,\mpipi)$. The size of this component was estimated using the \pythia6\ event generator~\cite{Sjostrand:2003wg} tuned to \hermes\ data~\cite{Liebing:2004us}. The strange contribution was neglected, while isospin and charge-conjugation symmetry implies that both $D_{1,q}^{p}(z,\mpipi)$ as well as $D_{1,q}(z,\mpipi)$ have identical values for $q=u,\bar{u},d,\bar{d}$. Varying the \pythia6\ estimate by  20\% does not significantly change the systematic uncertainty assigned to $a$.  The presented values for $a$ are the central values in the ranges of $a$ obtained by varying $b$ between its lower and upper bounds, while the ``b-scan'' uncertainty is taken to be the standard deviation.

%%%%%%%%%%%%%  RESULTS  %%%%%%%%%%%%%%%%

The values of the amplitudes $A_{U\perp}^{\sin(\phi_{R \perp} + \phi_S)\sin \theta}$ extracted as functions of $M_{\pi\pi}$, $x$, and $z$, are shown in Fig.~\ref{fig:asymmetries} and reported in 
Table~\ref{table:summary}. 
They are positive over the entire range of all three variables. The
reduced-$\chi^2$ values for the fits to the data set are in the range 0.64--1.38. 
The measured asymmetry is based on events integrated 
over $\boldsymbol{P_{h \perp}}$ (within the acceptance), 
which considerably simplifies an eventual
extraction of $h_1^q$ and $H_{1,q}^{\newangle,sp}$,  
since in this case $h_1^qH_{1,q}^{\newangle,sp}$  appears in the expression for the
 modulation amplitudes as a simple product (see Eq.~\eqref{eq:A_UT_theory}) instead of in
a  convolution integral over transverse momentum.   

\FIGURE{
\includegraphics[width=1.0\textwidth]{results.epsi}
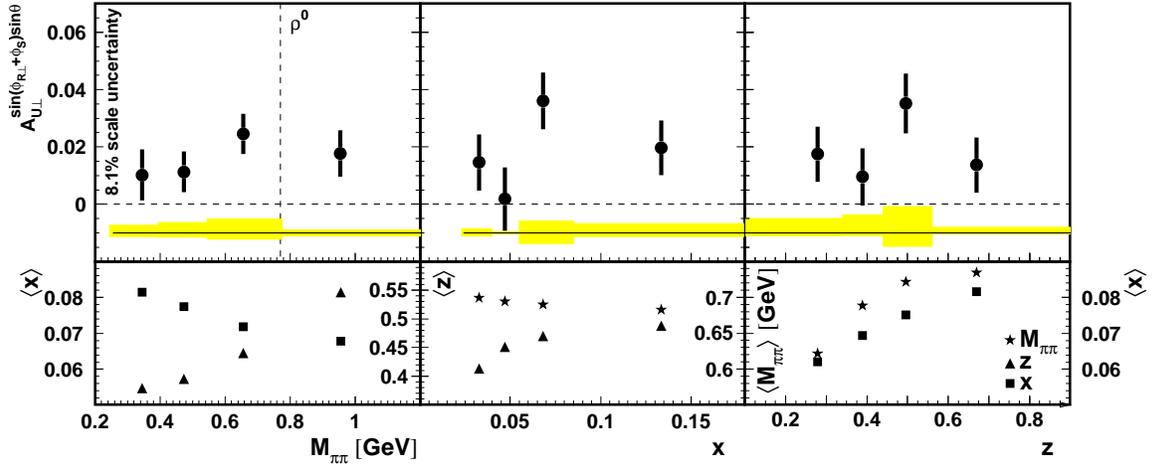
\caption{The top panels show $A_{U\perp}^{\sin(\phi_{R \perp} + \phi_S)\sin \theta}$ 
	versus \mpipi, $x$, and $z$.  
  	The bottom panels show the average values of the variables that were
        	integrated over. For the dependence on $x$ and $z$, $M_{\pi\pi}$ was
        	constrained to the range  $0.5$\,\,GeV $<M_{\pi\pi}<1.0$\,\,GeV,
	where the signal is expected to be largest. 
	The error bars show the statistical uncertainty. A scale 
	uncertainty of 8.1\% arises from the uncertainty in the target polarization. 
	Other contributions to the systematic uncertainty are summed in quadrature
	and represented by the asymmetric error band. 
	}
\label{fig:asymmetries}
}

The value of $A_{U\perp}^{\sin(\phi_{R \perp} + \phi_S)\sin \theta}$ extracted from
events summed over the experimental acceptance is 
$A_{U\perp}^{\sin(\phi_{R \perp} + \phi_S)\sin \theta} 
= 0.018  \pm 0.005_{\rm stat} \pm  0.002_{\rm b-scan}$, 
with an additional 8.1\% scale uncertainty coming from the uncertainty in the 
determination of the target polarization.  
As discussed below, 
acceptance effects were found to lead to an underestimate 
of the true value of the  modulation amplitude by up to 20\%.
For this result, the ranges selected in $x$ and $M_{\pi\pi}$
are $0.023 < x < 0.4$ and $0.5$\,\,GeV $< M_{\pi\pi} < 1.0$\,\,GeV. The
mean values of the kinematic variables are $\av{x} = 0.07$, $\av{y} = 0.64$,
$\av{Q^2} = 2.35$\,\,GeV$^2$, $\av{z} = 0.43$, and
$\av{|\boldsymbol{P_{h\perp}}|} = 0.42$\,\,GeV.

%%%%%%%%%%  SYSTEMATICS   %%%%%%%%%%%%%%%%

The  modulation amplitudes extracted are not influenced by the addition in the fit of terms of the form   
$\sin\phi_S$ (which appears at subleading twist in the polarized cross section 
$\sigma_{UT}$), or of the form $\cos\phi_{R \perp}\sin\theta$ (which appears
at subleading twist in the unpolarized cross section $\sigma_{UU}$). 
These angular combinations exhaust the possibilities up to subleading twist.
In order to eliminate effects of the natural polarization of the \hera\ lepton beam, data 
with both beam-helicity states were combined. The resulting net beam polarization is
$-0.020\pm0.001$. 
The influence of this small but nonzero net polarization on the amplitude extracted 
was shown to be negligible by analyzing separately the data of the two beam-helicity states.
There is also no influence from the addition to the fit of a constant term, the latter  being
consistent with zero. Identical results were obtained using an unbinned
maximum-likelihood fit.

Tracking corrections that are applied for the deflections of the scattered particles
caused by the vertical 0.3\,\,T target holding field have also  a
negligible effect on the extracted asymmetries.

The fully differential asymmetry depends on nine kinematic variables: $x$,
$y$, $z$, $\phi_{R \perp}$, $\phi_S$, and $\theta$, $M_{\pi\pi}$, 
and $\boldsymbol{P_{h \perp}}$ 
($\de^2 \boldsymbol{P_{h \perp}} = |\boldsymbol{P_{h \perp}}| \de |\boldsymbol{P_{h \perp}}| \de \phi_h$). 
Due to the limited statistical precision, it is not possible to measure the 
asymmetry $A_{U\perp}$ fully differential in all relevant variables. Combined
with the fact that the \hermes\ spectrometer does not have a full 4$\pi$
acceptance, this implies that the measured number of events is always
convolved with the experimental acceptance $\epsilon$, e.g., 
\begin{eqnarray}
N^{\uparrow(\downarrow)}(\phi_{R\perp},\phi_S,\theta,M_{\pi\pi}) &\propto& 
\int \! \de x\de y \de z \de^2\boldsymbol{P_{h \perp}}\,
    \epsilon(x,y,z,\boldsymbol{P_{h \perp}},\phi_{R
    \perp},\phi_S,\theta,M_{\pi\pi}) \ \times \nonumber \\
& & \times  \ \ \sigma_{U\uparrow(\downarrow)}(x,y,z,\boldsymbol{P_{h\perp}},\phi_{R \perp},\phi_S,\theta,M_{\pi\pi}),  
\end{eqnarray}
such that $\epsilon$ does not necessarily drop out of the expression for the
asymmetry (Eq.~\eqref{eq:asymmetry})\footnote{Note that, experimentally, the asymmetry 
itself is never integrated directly over any variables: always the numerator and denominator of the 
asymmetry are integrated separately.}.
Some effects of the acceptance can be easily dealt with if the predicted asymmetry amplitude is linearly dependent 
on all variables in the range over which they are integrated.  
In that case, the measured amplitudes are equal to the true amplitudes evaluated at the average values of these variables.
However, all models predict a highly nonlinear behavior of the amplitude as a function of the invariant mass $M_{\pi\pi}$.
Moreover, when the integration of the cross section over 
\(\boldsymbol{P_{h \perp}}\)  is incomplete because of the geometrical acceptance,  
other terms in  the \(\boldsymbol{P_{h \perp}}\)-unintegrated cross section~\cite{Radici:2001na,Bacchetta:2002ux} might contribute to the extracted amplitudes.

Therefore, a systematic uncertainty was estimated based on a Monte Carlo study, which is explained in more detail in the Appendix. In particular, two possible sources of systematic uncertainties have been examined: the difference in the  modulation amplitude of interest extracted as done for real data in the experimental acceptance and similarly in \(4\pi\) acceptance, and a possible false asymmetry originating from other terms appearing through incomplete integration over \(\boldsymbol{P_{h \perp}}\).

The largest effect was seen when comparing the amplitudes in  \(4\pi\) 
and in the experimental acceptance. 
The Monte Carlo simulation used a particular choice for transversity and for each of the dihadron fragmentation functions, which results in a reasonable description of the kinematic dependences of the measured amplitudes (cf.~Figs.~\ref{fig:MCamplitudes} and \ref{fig:asymmetries}). The amplitudes extracted in the experimental acceptance were found to be underestimated by 
up to 43\% for certain values of $z$ when compared to amplitudes extracted in \(4\pi\) coverage.
The effect was negligible for all $x$ bins when integrating over $z$, and about 21\% when integrated over the whole kinematic range.
No other models for the dihadron functions involved, suitable for this simulation, are presently available.
This systematic uncertainty estimate applies only when interpreting the results as values based on separate integration of numerator and denominator of the asymmetry over the relevant ranges of all kinematic variables. This choice was necessitated by the strong model-dependence of the acceptance effects when not integrating over \mpipi.

The incomplete integration over \(\boldsymbol{P_{h \perp}}\) was found to have only a small
influence on the extracted amplitudes due to possible terms in the \(\boldsymbol{P_{h \perp}}\)-unintegrated  cross section~\cite{Radici:2001na,Bacchetta:2002ux}. In view of the large 
uncertainties above, it can be neglected.

\TABLE{
\caption{The extracted  modulation amplitudes with statistical uncertainty, the systematic uncertainties 
	      	arising from the scan of $b$ in the fits and from extracting the amplitudes in the 
	      	experimental acceptance as described in the text. A further  8.1\% scale uncertainty 
	      	from the target polarization is not listed. 
		In addition, the bin boundaries are given in the various $M_{\pi\pi}$-, $x$-, and 
		$z$-bins, respectively, as well as the reduced-$\chi^2$ values of the fits.  
		Note that for both the $x$ and $z$ dependences, the lower and upper limits on 
		$M_{\pi\pi}$ are 0.5~GeV and 1~GeV, respectively. 
		The acceptance effect in the last row is not an average over those values for 
		$x$ or $z$ bins because the bin weighting for the amplitudes in 4$\pi$ differs from
		those for experimental acceptance.
		\vspace*{1.3mm}}\label{table:summary}  
\begin{tabular*}{\textwidth}{c|c|c}
\hline\hline\\[-4.5mm]
\hspace*{1.cm} bin boundaries 		\hspace*{1.2cm}						& 
\hspace*{2cm} \( A_{U\perp}^{\sin(\phi_{R \perp} + \phi_S)\sin \theta}\)  	\hspace*{2cm}	& 
~reduced $\chi^2$	\\[0.5mm]
\hline
0.25~GeV$<M_{\pi\pi}<$0.40~GeV 	& \( 0.010  \pm 0.009_{\rm stat} \pm  0.001_{\rm b-scan} + 0.002_{\rm acc}\)  &  0.70  \\
0.40~GeV$<M_{\pi\pi}<$0.55~GeV 	& \( 0.012  \pm 0.007_{\rm stat} \pm  0.001_{\rm b-scan} + 0.003_{\rm acc}\)  &  1.32  \\
0.55~GeV$<M_{\pi\pi}<$0.77~GeV	& \( 0.024  \pm 0.007_{\rm stat} \pm  0.002_{\rm b-scan} + 0.004_{\rm acc}\)  &  0.85  \\
0.77~GeV$<M_{\pi\pi}<$2.00~GeV	& \( 0.019  \pm 0.008_{\rm stat} \pm  0.001_{\rm b-scan} + 0.000_{\rm acc}\)  &  0.96  \\
\hline
0.023$<x<$0.040 & \( 0.015 \pm  0.010_{\rm stat} \pm  0.001_{\rm b-scan} + 0.001_{\rm acc}\)  &  0.88  \\
0.040$<x<$0.055 & \( 0.002 \pm  0.011_{\rm stat} \pm  0.001_{\rm b-scan} + 0.000_{\rm acc}\)  &  1.03  \\
0.055$<x<$0.085 & \( 0.035 \pm  0.010_{\rm stat} \pm  0.004_{\rm b-scan} + 0.002_{\rm acc}\)  &  1.38  \\
0.085$<x<$0.400 & \( 0.020 \pm  0.010_{\rm stat} \pm  0.001_{\rm b-scan} + 0.003_{\rm acc}\)  &  0.94  \\
\hline
0.000$<z<$0.340 & \( 0.018 \pm  0.010_{\rm stat} \pm  0.001_{\rm b-scan} + 0.005_{\rm acc}\)  &  1.04  \\
0.340$<z<$0.440 & \( 0.010 \pm  0.010_{\rm stat} \pm  0.001_{\rm b-scan} + 0.006_{\rm acc}\)  &  0.64  \\
0.440$<z<$0.560 & \( 0.036 \pm  0.010_{\rm stat} \pm  0.005_{\rm b-scan} + 0.008_{\rm acc}\)  &  1.04  \\
0.560$<z<$1.000 & \( 0.012 \pm  0.009_{\rm stat} \pm  0.001_{\rm b-scan} + 0.002_{\rm acc}\)  &  0.84  \\
\hline
0.5~GeV$<M_{\pi\pi}<$1.0~GeV 	&  & \\
0.023$<x<$0.400 & \(0.018  \pm 0.005_{\rm stat} \pm  0.002_{\rm b-scan} + 0.004_{\rm acc}\) & 0.87 \\
0.0$<z<$1.0					&  & \\
\hline\hline
\end{tabular*}
}

The interpretation of the amplitudes extracted can, in principle, be complicated by the experimental 
condition that the target polarization is transverse to the beam axis instead of  transverse to the virtual-photon direction. These {\it beam-axis} asymmetries can receive contributions not only from the transverse component of the nucleon spin with respect to the virtual-photon
direction but also from a small longitudinal component proportional to \(\sin\theta_{\gamma^*}\), where \(\theta_{\gamma^*}\) is the angle between the directions of the virtual photon and the incoming lepton beam. Such a contribution to the amplitude presented here can occur only when a \(\sin\phi_{R\perp}\) amplitude exists in the corresponding asymmetry \(A_{UL}\), i.e., the {\it photon-axis} asymmetry in dihadron lepto-production with an unpolarized beam on a longitudinally polarized 
target~\cite{Diehl:2004}. Such an amplitude exists at subleading twist~\cite{Bacchetta:2003vn}, but was measured to be small for  pairs of unidentified hadrons~\cite{vanderNat:2005aq}. 
In addition, \(\langle \sin\theta_{\gamma^*}\rangle\) is typically less than 
0.09~\cite{vanderNat:2005aq}, leading to an insignificant difference in the presented amplitude for lepton-axis and photon-axis asymmetries.

Besides this contribution, no other twist-3 effects are present in the measured amplitude. 
Modifications due to even higher twist and NLO effects
are unknown for dihadron production in DIS. However, the dominant NLO contribution to the 
``longitudinal'' cross section \(\sigma_L\) is known to be up to 30\%
for the unpolarized {\it inclusive} DIS cross section in these kinematic 
conditions~\cite{Whitlow:1990gk}.

%%%%%%%%%%%  INTERPRETATION  %%%%%%%%%%%%%

Since the fragmentation functions $H_{1,q}^{\newangle,sp}$ require the
interference between $s$ and $p$ waves, it is supposed to be sizeable in
the regions where spin-1 resonances are present, assuming the rest of the 
spectrum to be in an $s$ wave. As can be seen 
in Fig.~\ref{fig:distributions}, in the invariant-mass range explored in this
paper the $\rho^0$ and $\omega$ resonances are present and give large
contributions to the spectrum. The available theoretical
models indicate that $H_{1,q}^{\newangle,sp}$
should be maximal in the vicinity of the $\rho^0$
mass~\cite{Jaffe:1998hf,Radici:2001na,Bacchetta:2006un}. 

\FIGURE{
\includegraphics[width=10.cm]{InvMassSpectra.epsi}
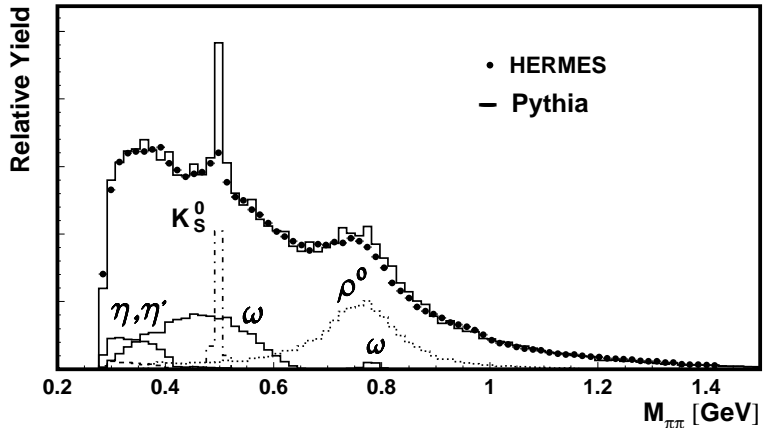
\caption{Yield distribution in the invariant mass of the $\pi^+\pi^-$ pairs for the
	experimental data compared to a \pythia6\ Monte Carlo simulation. 
	Both distributions are normalized to unity. The main resonances contributing 
	to the simulated spectrum are shown separately.}
\label{fig:distributions}
}

Being naive-T-odd, the fragmentation function requires the
interference between scattering amplitudes with different phases. The model of
Ref.~\cite{Jaffe:1998hf} considers the interference between the $\rho^0$ and
the $\sigma$ resonance, as measured in $\pi^+ \pi^-$ scattering,
predicting a sign change of the fragmentation function close to the $\rho^0$
mass.   
The models of Refs.~\cite{Radici:2001na,Bacchetta:2006un} neglect the
contributions from the $\sigma$ resonance and assume the $s$-wave amplitude 
of the spectrum to be real. Thus, the fragmentation function turns out to be
almost proportional to the imaginary part of the $\rho^0$ resonance, i.e., a
Breit--Wigner shape peaked at the  $\rho^0$ mass. 
In Ref.~\cite{Bacchetta:2006un}, the imaginary part of the $\omega$ resonance 
is also taken into account, giving rise to an additional contribution to the
fragmentation function in the region around $M_{\pi \pi} \approx 0.5$~GeV.

The $M_{\pi\pi}$ dependence of the measured  modulation amplitude shows no
sign change at the $\rho^0$ mass, contrary to the prediction in
Ref.~\cite{Jaffe:1998hf}. This leads to the conclusion that $\rho$-$\sigma$
interference is not the dominant contribution to the fragmentation function
$H_1^{\newangle,sp}$, and that in general interference patterns observed in
semi-inclusive $\pi^+ \pi^-$ production are different from those observed in 
$\pi^+ \pi^-$ scattering. 
The dependences on $M_{\pi\pi}$ and $z$ of the model calculations of
Ref.~\cite{Bacchetta:2006un} (see also \cite{She:2007ht}), one of which is 
reproduced in Fig.~\ref{fig:MCamplitudes}, are not inconsistent in shape with the 
present data. However, the predictions are at least a factor of two larger.

%%%%%%%%%%%%   Summary & Outlook   %%%%%%%%%%%%%%%

In summary, a measurement of $A_{U\perp}^{\sin(\phi_{R \perp} + \phi_S)\sin \theta}$
of the transverse-target-spin asymmetry in the
lepto-production of $\pi^+\pi^-$ pairs has provided the first evidence that
a naive-T-odd chiral-odd dihadron fragmentation function $H_{1,q}^{\newangle}$ 
and in particular $H_{1,q}^{\newangle,sp}$ is nonzero.  
The average value of the amplitude is
$A_{U\perp}^{\sin(\phi_{R \perp} + \phi_S)\sin \theta} 
= 0.018  \pm 0.005_{\rm stat} 
\pm  0.002_{\rm b-scan}  
+  0.004_{\rm acc} $,
with an additional 8.1\% scale uncertainty.
The amplitude is positive in the whole range 
in the  invariant mass of the  $\pi^+\pi^-$ pairs, in contrast to a previous
expectation~\cite{Jaffe:1998hf} of a sign change around the mass
of the \(\rho^0\)  meson.  
Possibly the most striking aspect of the
reported results is the relatively large size of an asymmetry 
caused by a complicated interference effect.

A mechanism analogous to the one investigated in this paper offers perhaps
the most promising way to access transversity in $pp$ collisions at \rhic. Our
results show for the first time that this mechanism can indeed give a sizeable
signal.
The {\belle} collaboration can extract dihadron fragmentation functions from 
their $e^+e^-$ data. Such results could then be combined with DIS and $pp$ 
data to extract transversity in the proton.

%%%%%%%%%%%%   ACKNOWLEDGMENTS   %%%%%%%%%%%%%%%

\acknowledgments

We gratefully acknowledge the \desy\ management for its support and the staff
at \desy\ and the collaborating institutions for their significant effort.
This work was supported by the FWO-Flanders and IWT, Belgium;
the Natural Sciences and Engineering Research Council of Canada;
the National Natural Science Foundation of China;
the Alexander von Humboldt Stiftung;
the German Bundesministerium f\"ur Bildung und Forschung (BMBF);
the Deutsche Forschungsgemeinschaft (DFG);
the Italian Istituto Nazionale di Fisica Nucleare (INFN);
the MEXT, JSPS, and COE21 of Japan;
the Dutch Foundation for Fundamenteel Onderzoek der Materie (FOM);
the U. K. Engineering and Physical Sciences Research Council, the
Particle Physics and Astronomy Research Council and the
Scottish Universities Physics Alliance;
the U. S. Department of Energy (DOE) and the National Science Foundation (NSF);
the Russian Academy of Science and the Russian Federal Agency for 
Science and Innovations;
the Ministry of Trade and Economical Development and the Ministry
of Education and Science of Armenia;
and the European Community-Research Infrastructure Activity under the
FP6 ''Structuring the European Research Area'' program
(Hadron Physics, contract number RII3-CT-2004-506078).

%%%%%%%%%%%%%    APPENDIX     %%%%%%%%%%%%%%%%

\appendix

\section{Description of the Monte Carlo Study}

The starting point of the acceptance studies was a \pythia6\ Monte Carlo 
simulation~\cite{Sjostrand:2003wg}, 
which does not have any processes related to transverse target polarization.
Specifically, a version of \pythia6\ was used where the relevant cross sections
 were tuned to \hermes\ data~\cite{Liebing:2004us}.
The target-polarization dependence was introduced by randomly assigning spin states to events
with a probability according to an expression for $A_{UT}$ as a function of the various kinematic
variables.

In the first study, only the  modulation amplitude of interest was implemented 
in order to assess the effects of the acceptance on it.
For the dihadron fragmentation functions $D_{1,q}(z,M_{\pi\pi})$ and 
$H_{1,q}^{\newangle,sp}(z,M_{\pi\pi})$, the models  of Ref.~\cite{Bacchetta:2006un} 
were implemented. 
For the distribution functions $f_{1}^{q}(x)$ and $h_{1}^{q}(x)$, 
parameterizations were taken from
Ref.~\cite{Gluck:1998xa} and from Ref.~\cite{Schweitzer:2001sr}, respectively. 
No additional dependence on transverse momentum was introduced, i.e., it was assumed that any dependence on transverse momentum of the products of polarized and unpolarized distribution and fragmentation functions cancels in the asymmetry.

Modulation amplitudes were then extracted 
in a fit\footnote{For this study it was assumed that the acceptance
in $\theta$ is complete, i.e., no contribution from $b$ was taken into account in 
Eq.~\eqref{eq:fit_function}.} to both the data in \(4\pi\) and the {\hermes} experimental acceptance, 
where the latter was simulated with a parameterization of the spectrometer 
performance based on \geant3.
The shape of the yield distributions in all nine kinematic variables in the experimental acceptance can be found in Ref.~\cite{vanderNat:2007zz}. 
As shown in Fig.~\ref{fig:MCamplitudes}, the acceptance effect can be quite large: the  modulation amplitudes extracted in the experimental acceptance are  underestimated by up to 25\%
in certain $M_{\pi\pi}$ bins and by up to 43\% for certain $z$ bins
when compared to amplitudes extracted in \(4\pi\) coverage. 
The effect was negligible in all $x$ bins. 
Apparent is the discrepancy in the average values of $x$ for 4$\pi$ and the 
experimental acceptances, where a strong dependence of the asymmetry on $x$, which is driven by the increase of transversity with $x$ in the range considered, leads to 
the observed underestimates in the amplitudes extracted when integrated over $x$.

\FIGURE{
\includegraphics[width=1.0\textwidth]{4piHermesComparison.epsi}
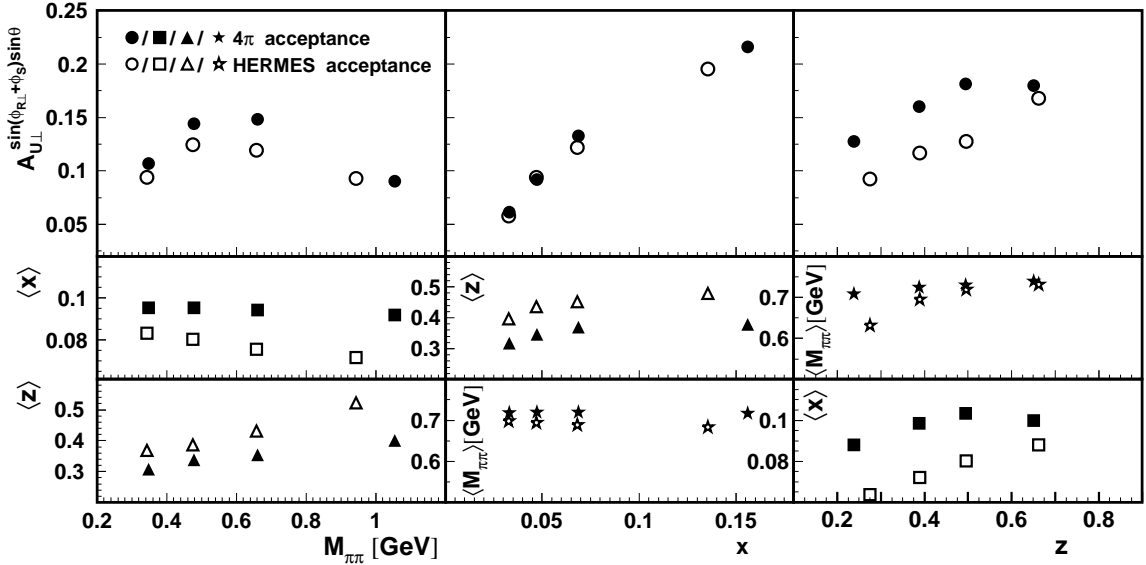
\caption{The top panels show $A_{U\perp}^{\sin(\phi_{R \perp} + \phi_S)\sin \theta}$ 
	versus \mpipi, $x$, and $z$ for Monte-Carlo data extracted both in 4$\pi$ 
	and experimental acceptance.  
  	The bottom panels show the average values of the variables that were
        	integrated over. For the dependence on $x$ and $z$, $M_{\pi\pi}$ was
        	constrained to the range  $0.5$\,\,GeV $<M_{\pi\pi}<1.0$\,\,GeV.
	The systematic uncertainties assigned to the amplitudes extracted from real
	data (listed in Table~\ref{table:summary}) are obtained from the
	differences between the above amplitudes in the experimental acceptance 
	as compared to 4$\pi$. These differences were scaled by the ratio of the 
	average reconstructed amplitudes obtained from HERMES data and from the 
	Monte Carlo data in order to accommodate the larger magnitude of the model prediction.}
\label{fig:MCamplitudes}
}

A second study dealt with contributions from contaminating  modulation amplitudes appearing through the
incomplete integration over  \(\boldsymbol{P_{h \perp}}\).
The experimental acceptance has a strong dependence on $\phi_h$, the azimuthal angle of
$P_{h}$ around the virtual-photon direction,
with the consequence that the extracted amplitude $a$ in Eq.~\eqref{eq:fit_function} does not necessarily correspond to Eq.~\eqref{eq:A_UT_theory}.  The fully differential
$\phi_h$-dependent cross section~\cite{Radici:2001na,Bacchetta:2002ux} 
contains many terms, which if nonzero and if the integral over $\phi_h$ is incomplete, can give unwanted contributions to the  modulation amplitude.

In principle, these terms could be taken into account in the fit (Eq.~\eqref{eq:fit_function}), 
but this is difficult with the current statistical precision of this measurement, 
as it would require, e.g., 3-dimensional binning, i.e., an additional binning in \(\phi_h\).
To study the influence on the amplitude of interest, model predictions for the size and dependences 
of all these $\phi_h$-dependent terms are necessary. However, no such information exists, i.e., most of the distribution and fragmentation functions involved are as yet completely unknown. 
In order to estimate a systematic uncertainty, a very general model was used
for these terms, varying their size and dependences. The averages of the effect on the extracted 
value of $A_{U\perp}^{\sin(\phi_{R \perp} +\phi_S)\sin \theta}$ were then used to estimate a contribution to the systematic uncertainty.

Target spin states were again assigned to semi-inclusive events from a \pythia6\ Monte Carlo simulation 
according to a model for the asymmetry $A_{UT}$, but now including all $\phi_h$-dependent terms. 
For the distribution functions $f_1^q(x)$, $h_1^q(x)$ and for the fragmentation functions 
$D_{1,q}(M_{\pi\pi},z)$ and $H_{1,q}^{\newangle,sp}(M_{\pi\pi},z)$ the same models were 
used as before. 
For the transverse-momentum dependence of all distribution and fragmentation functions appearing 
in $A_{UT}$, a Gaussian Ansatz was used:
\begin{eqnarray}
f_{1}(x,\mb{p_T^\textrm{2}}) &=& \frac{1}{\pi \langle \mb{p_T^\textrm{2}} \rangle} e^{-\frac{\mb{p_T^\textrm{2}}}{\langle \mb{p_T^\textrm{2}} \rangle } }
f_{1}(x)\mathrm{,}
\label{eq:gauss_DF} \\
D_1(z,M_{\pi\pi},\cos\theta,\mb{k_T^\textrm{2}},\mb{k_T}\cdot \mb{R_T}) &=& 
\frac{1}{z^2\pi \langle \mb{k_T^\textrm{2}} \rangle} e^{-\frac{\mb{k_T^\textrm{2}}}{\langle \mb{k_T^\textrm{2}} \rangle} } D_1(z,M_{\pi\pi},\cos\theta).
\label{eq:gauss_FF}
\end{eqnarray}
with $\mb{p_T}$ ($\mb{k_T}$) being the initial- (final-/fragmenting-) quark's momentum component that is transverse to the initial- (final-) hadron's momentum direction.
The same $\mb{p_T^\textrm{2}}$ and $\mb{k_T^\textrm{2}}$ dependences were used for all other distribution and fragmentation functions. The actual values of $\mb{p_T^\textrm{2}}$ and 
$\mb{k_T^\textrm{2}}$ are irrelevant as they are absorbed in the $C_N$ in Eq.~\ref{eq:phih_amplitudes}.

The $\phi_h$-dependent terms were implemented 
such that the corresponding azimuthal amplitudes
$A_{UT}^{\sin(a \phi_h + b \phi_{R \perp} + c \phi_S + \frac{d}{2} \pi )}$ 
depend on $x$, $z$, and $\boldsymbol{P_{h \perp}}$ according to
\begin{eqnarray}
\frac{1}{2} A_{UT}^{\sin(a \phi_h + b \phi_{R \perp} + c \phi_S + \frac{d}{2} \pi )} 
&\equiv& 
\frac{\int \!\! \de \phi_h \de \phi_{R \perp} \de \phi_S 
\sin(a \phi_h + b \phi_{R \perp} + c \phi_S + \frac{d}{2} \pi) \de^9 
\sigma_{UT}}{\int \!\! \de \phi_h \de \phi_{R \perp} \de \phi_S \de^9 
\sigma_{UU}} \nonumber \\
&=& C_N \ z^{\alpha_N} \ x^{\beta_N} \ f_N(|\boldsymbol{P_{h \perp}}|) ,
\label{eq:phih_amplitudes}
\end{eqnarray}
with $N$ identifying the various possible terms in the full polarized 
cross section~\cite{Radici:2001na,Bacchetta:2002ux}, 
$C_N$ a constant scaling factor, $\alpha_N$, $\beta_N \in [0.1,3]$ 
and $a,b,c$ and $d$ are either zero or integers depending on $N$. 
The interval $[0.1,3]$ is based on typical parameterizations of the parton 
distributions $f_1^q$ and the single-hadron fragmentation function $D_{1,q}(z)$.
Similarly, azimuthal amplitudes $A_{UU}^{\sin(a \phi_h + b \phi_{R \perp} + c 
\phi_S + \frac{d}{2} \pi )}$ were introduced for the $\phi_h$-dependent 
parts in the unpolarized cross section.
Apart from the fact that all these different  modulation amplitudes of the polarized and 
unpolarized cross section increase nonlinearly with increasing 
$x$ and $z$, the choices for $\alpha_N$ and $\beta_N$ are quite
arbitrary, but were found not to influence the final conclusions.
Starting from the expressions for the convolution integrals in the involved cross 
sections~\cite{Radici:2001na,Bacchetta:2002ux} and  
using the Gaussian Ans\"atze Eqs.~(\ref{eq:gauss_DF},\ref{eq:gauss_FF}) 
for the $\mb{p_T^\textrm{2}}$ and 
$\mb{k_T^\textrm{2}}$ dependence of the distribution and fragmentation 
functions, the dependences
$f_N(|\boldsymbol{P_{h \perp}}|)$ of the  modulation amplitudes on  $|\boldsymbol{P_{h \perp}}|$
were derived~\cite{vanderNat:2007zz}.\footnote{Due 
to the fact that no $\mb{k_T}\cdot \mb{R_T}$ dependence is 
taken into account for the fragmentation functions,
about half of the $\phi_h$-dependent terms drop out of the complete 
expression for the polarized cross section.}

The values of the scaling factors $C_N$ in Eq.~\ref{eq:phih_amplitudes} 
were derived from the averaged  modulation amplitudes, which were randomly chosen
in the range $[-0.1,0.1]$, i.e.,
\begin{equation}
\frac{\int A_{UU/T}^{\sin(a \phi_h + b \phi_{R \perp} + c \phi_S + 
\frac{d}{2} \pi)} \de^9 \sigma_{UU}}{\int \de^9 \sigma_{UU}} \in 
[-0.1,0.1],
\label{eq:R_integral}
\end{equation}
where the integral is performed over all nine dimensions and integration 
ranges were used corresponding to those used in the analysis.
Each resulting parameterization of $A_{UT}$ had to satisfy the positivity 
limit $|A_{UT}|<1$.

To estimate the systematic uncertainty, the amplitude 
$A_{U\perp}^{\sin(\phi_{R \perp} + \phi_S)\sin \theta}$ was extracted 
1000 times from the same \pythia6\ dataset, similar in size to the real data, but 
each time with spin states randomly chosen according to their probability 
calculated from randomly chosen values 
for $\alpha_N$, $\beta_N$, and $C_N$ for each of the $\phi_h$-dependent terms. 
The distribution obtained in the extracted amplitudes 
$A_{U\perp}^{\sin(\phi_{R \perp} + \phi_S)\sin \theta}$ was compared to a 
similarly obtained distribution,  but  which had only 
$A_{UT}^{\sin(\phi_{R \perp} + \phi_S)\sin \theta}$ implemented. 
On average the implementation of the $\phi_h$ dependence resulted in a distribution which 
has the same average value, but which is 10\% broader, 
independent of the $M_{\pi \pi}$, $z$ or $x$ bin considered.
Thus this effect is found to be small compared to the other effect of the acceptance described above.

%%%%%%%%%%%%%%  AUTHOR LIST  %%%%%%%%%%%%%%%%%%

\input{authorlist_JHEP}

%%%%%%%%%%%%%%  BIBLIOGRAPHY  %%%%%%%%%%%%%%%%%%

\bibliography{dihadron_JHEP}% Produces the bibliography via BibTeX.

\end{document}

%% file: authorlist_JHEP.tex
\section*{Author list}
\subsection*{The \hermes \ collaboration}

% authors
{%
%\begin{center}
\begin{flushleft} 
\bf
\hermesauthor[16]{A.~Airapetian},
\hermesauthor[27]{N.~Akopov},
\hermesauthor[27]{Z.~Akopov},
\hermesauthor[15]{A.~Andrus},
\hermesauthor[7]{E.C.~Aschenauer},
\hermesauthor[26]{W.~Augustyniak},
\hermesauthor[27]{R.~Avakian},
\hermesauthor[27]{A.~Avetissian},
\hermesauthor[11]{E.~Avetissian},
\hermesauthor[21]{A.~Bacchetta},
\hermesauthor[10]{L.~Barion},
\hermesauthor[19]{S.~Belostotski},
\hermesauthor[11]{N.~Bianchi},
\hermesauthor[18,25]{H.P.~Blok},
\hermesauthor[7]{H.~B\"ottcher},
\hermesauthor[10]{C.~Bonomo},
\hermesauthor[14]{A.~Borissov},
\hermesauthor[11]{A.~Borysenko},
\hermesauthor[]{A.~Br\"ull}\symbolfootnote[2]{Present address: 36 Mizzen Circle, Hampton, Virginia 23664, USA},
\hermesauthor[20]{V.~Bryzgalov},
\hermesauthor[10]{M.~Capiluppi},
\hermesauthor[11]{G.P.~Capitani},
\hermesauthor[22]{E.~Cisbani},
\hermesauthor[10]{G.~Ciullo},
\hermesauthor[10]{M.~Contalbrigo},
\hermesauthor[10]{P.F.~Dalpiaz},
\hermesauthor[16]{W.~Deconinck},
\hermesauthor[2]{R.~De~Leo},
\hermesauthor[18]{M.~Demey},
\hermesauthor[6,23]{L.~De~Nardo},
\hermesauthor[11]{E.~De~Sanctis},
\hermesauthor[17]{E.~Devitsin},
\hermesauthor[9]{M.~Diefenthaler},
\hermesauthor[11]{P.~Di~Nezza},
\hermesauthor[18]{J.~Dreschler},
\hermesauthor[13]{M.~D\"uren},
\hermesauthor[9]{M.~Ehrenfried},
\hermesauthor[1]{A.~Elalaoui-Moulay},
\hermesauthor[27]{G.~Elbakian},
\hermesauthor[5]{F.~Ellinghaus},
\hermesauthor[12]{U.~Elschenbroich},
\hermesauthor[18]{R.~Fabbri},
\hermesauthor[11]{A.~Fantoni},
\hermesauthor[23]{L.~Felawka},
\hermesauthor[22]{S.~Frullani},
\hermesauthor[11]{A.~Funel},
\hermesauthor[20]{G.~Gapienko},
\hermesauthor[20]{V.~Gapienko},
\hermesauthor[22]{F.~Garibaldi},
\hermesauthor[6,19,23]{G.~Gavrilov},
\hermesauthor[27]{V.~Gharibyan},
\hermesauthor[10]{F.~Giordano},
\hermesauthor[10]{O.~Grebeniouk},
\hermesauthor[7]{I.M.~Gregor},
\hermesauthor[18]{K.~Griffioen},
\hermesauthor[7]{H.~Guler},
\hermesauthor[11]{C.~Hadjidakis},
\hermesauthor[6]{M.~Hartig},
\hermesauthor[11]{D.~Hasch},
\hermesauthor[24]{T.~Hasegawa},
\hermesauthor[18,25]{W.H.A.~Hesselink},
\hermesauthor[14]{G.~Hill},
\hermesauthor[9]{A.~Hillenbrand},
\hermesauthor[13]{M.~Hoek},
\hermesauthor[6]{Y.~Holler},
\hermesauthor[12]{B.~Hommez},
\hermesauthor[7]{I.~Hristova},
\hermesauthor[8]{G.~Iarygin},
\hermesauthor[24]{Y.~Imazu},
\hermesauthor[20]{A.~Ivanilov},
\hermesauthor[19]{A.~Izotov},
\hermesauthor[1]{H.E.~Jackson},
\hermesauthor[19]{A.~Jgoun},
\hermesauthor[14]{R.~Kaiser},
\hermesauthor[13]{T.~Keri},
\hermesauthor[5]{E.~Kinney},
\hermesauthor[5,19]{A.~Kisselev},
\hermesauthor[24]{T.~Kobayashi},
\hermesauthor[7]{M.~Kopytin},
\hermesauthor[20]{V.~Korotkov},
\hermesauthor[17]{V.~Kozlov},
\hermesauthor[9]{B.~Krauss},
\hermesauthor[19]{P.~Kravchenko},
\hermesauthor[8]{V.G.~Krivokhijine},
\hermesauthor[2]{L.~Lagamba},
\hermesauthor[18]{L.~Lapik\'as},
\hermesauthor[10]{P.~Lenisa},
\hermesauthor[7]{P.~Liebing},
\hermesauthor[15]{L.A.~Linden-Levy},
\hermesauthor[16]{W.~Lorenzon},
\hermesauthor[13]{S.~Lu},
\hermesauthor[24]{X.~Lu},
\hermesauthor[3]{B.-Q.~Ma},
\hermesauthor[12]{B.~Maiheu},
\hermesauthor[15]{N.C.R.~Makins},
\hermesauthor[3]{Y.~Mao},
\hermesauthor[26]{B.~Marianski},
\hermesauthor[27]{H.~Marukyan},
\hermesauthor[18]{V.~Mexner},
\hermesauthor[23]{C.A.~Miller},
\hermesauthor[24]{Y.~Miyachi},
\hermesauthor[11]{V.~Muccifora},
\hermesauthor[14]{M.~Murray},
\hermesauthor[8]{A.~Nagaitsev},
\hermesauthor[2]{E.~Nappi},
\hermesauthor[19]{Y.~Naryshkin},
\hermesauthor[7]{M.~Negodaev},
\hermesauthor[7]{W.-D.~Nowak},
\hermesauthor[14]{A.~Osborne},
\hermesauthor[10]{L.L.~Pappalardo},
\hermesauthor[13]{R.~Perez-Benito},
\hermesauthor[9]{N.~Pickert},
\hermesauthor[9]{M.~Raithel},
\hermesauthor[9]{D.~Reggiani},
\hermesauthor[1]{P.E.~Reimer},
\hermesauthor[18]{A.~Reischl},
\hermesauthor[11]{A.R.~Reolon},
\hermesauthor[9]{C.~Riedl},
\hermesauthor[9]{K.~Rith},
\hermesauthor[6]{S.E.~Rock},
\hermesauthor[14]{G.~Rosner},
\hermesauthor[6]{A.~Rostomyan},
\hermesauthor[13]{L.~Rubacek},
\hermesauthor[15]{J.~Rubin},
\hermesauthor[12]{D.~Ryckbosch},
\hermesauthor[20]{Y.~Salomatin},
\hermesauthor[1,19]{I.~Sanjiev},
\hermesauthor[21]{A.~Sch\"afer},
\hermesauthor[24]{G.~Schnell},
\hermesauthor[6]{K.P.~Sch\"uler},
\hermesauthor[13]{B.~Seitz},
\hermesauthor[14]{C.~Shearer},
\hermesauthor[24]{T.-A.~Shibata},
\hermesauthor[8]{V.~Shutov},
\hermesauthor[10]{M.~Stancari},
\hermesauthor[10]{M.~Statera},
\hermesauthor[9]{E.~Steffens},
\hermesauthor[18]{J.J.M.~Steijger},
\hermesauthor[13]{H.~Stenzel},
\hermesauthor[7]{J.~Stewart},
\hermesauthor[9]{F.~Stinzing},
\hermesauthor[13]{J.~Streit},
\hermesauthor[9]{P.~Tait},
\hermesauthor[27]{S.~Taroian},
\hermesauthor[20]{B.~Tchuiko},
\hermesauthor[17]{A.~Terkulov},
\hermesauthor[26]{A.~Trzcinski},
\hermesauthor[12]{M.~Tytgat},
\hermesauthor[12]{A.~Vandenbroucke},
\hermesauthor[18]{P.B.~van~der~Nat},
\hermesauthor[18]{G.~van~der~Steenhoven},
\hermesauthor[12]{Y.~van~Haarlem},
\hermesauthor[12]{C.~van~Hulse},
\hermesauthor[19]{D.~Veretennikov},
\hermesauthor[19]{V.~Vikhrov},
\hermesauthor[9]{C.~Vogel},
\hermesauthor[3]{S.~Wang},
\hermesauthor[4]{Y.~Ye},
\hermesauthor[6]{Z.~Ye},
\hermesauthor[23]{S.~Yen},
\hermesauthor[9]{D.~Zeiler},
\hermesauthor[12]{B.~Zihlmann},
\hermesauthor[26]{P.~Zupranski}
\end{flushleft} 
%\end{center}
}
%-- HERMES Institutes
\bigskip
{\it
%\begin{center}
\begin{flushleft} 
\hermesinstitute[1]{Physics Division, Argonne National Laboratory, Argonne, Illinois 60439-4843, USA}
\hermesinstitute[2]{Istituto Nazionale di Fisica Nucleare, Sezione di Bari, 70124 Bari, Italy}
\hermesinstitute[3]{School of Physics, Peking University, Beijing 100871, China}
\hermesinstitute[4]{Department of Modern Physics, University of Science and Technology of China, Hefei, Anhui 230026, China}
\hermesinstitute[5]{Nuclear Physics Laboratory, University of Colorado, Boulder, Colorado 80309-0390, USA}
\hermesinstitute[6]{DESY, 22603 Hamburg, Germany}
\hermesinstitute[7]{DESY, 15738 Zeuthen, Germany}
\hermesinstitute[8]{Joint Institute for Nuclear Research, 141980 Dubna, Russia}
\hermesinstitute[9]{Physikalisches Institut, Universit\"at Erlangen-N\"urnberg, 91058 Erlangen, Germany}
\hermesinstitute[10]{Istituto Nazionale di Fisica Nucleare, Sezione di Ferrara and Dipartimento di Fisica, Universit\`a di Ferrara, 44100 Ferrara, Italy}
\hermesinstitute[11]{Istituto Nazionale di Fisica Nucleare, Laboratori Nazionali di Frascati, 00044 Frascati, Italy}
\hermesinstitute[12]{Department of Subatomic and Radiation Physics, University of Gent, 9000 Gent, Belgium}
\hermesinstitute[13]{Physikalisches Institut, Universit\"at Gie{\ss}en, 35392 Gie{\ss}en, Germany}
\hermesinstitute[14]{Department of Physics and Astronomy, University of Glasgow, Glasgow G12 8QQ, United Kingdom}
\hermesinstitute[15]{Department of Physics, University of Illinois, Urbana, Illinois 61801-3080, USA}
\hermesinstitute[16]{Randall Laboratory of Physics, University of Michigan, Ann Arbor, Michigan 48109-1040, USA }
\hermesinstitute[17]{Lebedev Physical Institute, 117924 Moscow, Russia}
\hermesinstitute[18]{National Institute for Subatomic Physics (Nikhef), 1009 DB Amsterdam, The Netherlands}
\hermesinstitute[19]{Petersburg Nuclear Physics Institute, St. Petersburg, Gatchina, 188350 Russia}
\hermesinstitute[20]{Institute for High Energy Physics, Protvino, Moscow region, 142281 Russia}
\hermesinstitute[21]{Institut f\"ur Theoretische Physik, Universit\"at Regensburg, 93040 Regensburg, Germany}
\hermesinstitute[22]{Istituto Nazionale di Fisica Nucleare, Sezione Roma 1, Gruppo Sanit\`a and Physics Laboratory, Istituto Superiore di Sanit\`a, 00161 Roma, Italy}
\hermesinstitute[23]{TRIUMF, Vancouver, British Columbia V6T 2A3, Canada}
\hermesinstitute[24]{Department of Physics, Tokyo Institute of Technology, Tokyo 152, Japan}
\hermesinstitute[25]{Department of Physics and Astronomy, Vrije Universiteit, 1081 HV Amsterdam, The Netherlands}
\hermesinstitute[26]{Andrzej Soltan Institute for Nuclear Studies, 00-689 Warsaw, Poland}
\hermesinstitute[27]{Yerevan Physics Institute, 375036 Yerevan, Armenia}
\end{flushleft} 
%\end{center}
}